\documentclass{article}
\usepackage{dahlin,enumitem,hyperref,testJD,natbib,multirow,rotating}           

\title{Particle Metropolis-Hastings using gradient and Hessian information\thanks{The final publication is available at Springer via: \newline \url{http://dx.doi.org/10.1007/s11222-014-9510-0}}}
\author{Johan Dahlin, Fredrik Lindsten and Thomas B.\ Sch\"{o}n%
\thanks{This work was supported by: Learning of complex dynamical systems (Contract number: 637-2014-466) and Probabilistic modeling of dynamical systems (Contract number: 621-2013-5524) and CADICS, a Linnaeus Center, all funded by the Swedish Research Council. JD is with the Division of Automatic Control, Link{\"o}ping University, Link{\"o}ping, Sweden. E-mail: \texttt{ johan.dahlin@liu.se}. FL is with the Department of Engineering, University of Cambridge, Cambridge, United Kingdom. E-mail: \texttt{fredrik.lindsten@eng.cam.ac.uk}. TS is with Division of Systems and Control, Uppsala University, Uppsala, Sweden. E-mail: \texttt{thomas.schon@it.uu.se}.}%
}

\begin{document}
\maketitle
\thispagestyle{empty}
\pagestyle{empty}

\begin{abstract}
Particle Metropolis-Hastings (PMH) allows for Bayesian parameter inference in nonlinear state space models by combining Markov chain Monte Carlo (MCMC) and particle filtering. The latter is used to estimate the intractable likelihood. In its original formulation, PMH makes use of a marginal MCMC proposal for the parameters, typically a Gaussian random walk.  However, this can lead to a poor exploration of the parameter space and an inefficient use of the generated particles.  

We propose a number of alternative versions of PMH that incorporate gradient and Hessian information about the posterior into the proposal. This information is more or less obtained as a byproduct of the likelihood estimation. Indeed, we show how to estimate the required information using a fixed-lag particle smoother, with a computational cost growing linearly in the number of particles. We conclude that the proposed methods can: (i) decrease the length of the burn-in phase, (ii) increase the mixing of the Markov chain at the stationary phase, and (iii) make the proposal distribution scale invariant which simplifies tuning.
\end{abstract}

\clearpage
\section{Introduction}
We are interested in Bayesian parameter inference in nonlinear state space models (SSM) of the form
\begin{align}
	x_{t}|x_{t-1}  \sim  f_{\theta}(x_{t}|x_{t-1}), \quad
	y_{t}|x_t      \sim  g_{\theta}(y_{t}|x_t),
\label{eq:SSM}
\end{align}
where the latent states and the measurements are denoted by $\mathbf{x} = x_{0:T} \triangleq \{x_t\}_{t=0}^T$ and $\mathbf{y} = \measurementsdef$, respectively. Here, $f_{\theta}(\cdot)$ and $g_{\theta}(\cdot)$ denote the transition and observation kernels, respectively, parametrised by the unknown static parameter vector $\theta \in \Theta \subset \mathbb{R}^d$. The initial state is distributed according to some distribution $\mu(x_0)$ which,
for notational simplicity, is assumed to be independent of $\theta$.

The aim of Bayesian parameter inference (in SSMs) is to compute the \textit{parameter posterior distribution}
\begin{align}
	p(\theta | \mathbf{y}) = \frac{ p_{\theta}(\mathbf{y}) p(\theta)}{p(\mathbf{y})},
	\label{eq:parameterPosterior}
\end{align}
where $p(\theta)$ denotes the prior of $\theta$ and $p_{\theta}(\mathbf{y})$ denotes the likelihood, which for an SSM can be expressed as
\begin{align}
	p_{\theta}(\mathbf{y}) = p_{\theta}(y_1) \prod_{t=2}^T p_{\theta}(y_t|y_{1:t-1}).
	\label{eq:likeFunc}
\end{align}
The one-step ahead predictor $p_{\theta}(y_t|y_{1:t-1})$, and thus also the likelihood function, is in general not analytically tractable. However, unbiased estimators of the likelihood can be constructed using sequential Monte Carlo (SMC) \citep{DoucetJohansen2011,DelMoral2004} and these can be used as \textit{plug-in estimators}. This is especially useful in the Metropolis-Hastings (MH) algorithm that can be used for estimating the parameter posterior in \eqref{eq:parameterPosterior}.

This combination of MH and SMC is known as the particle Metropolis-Hastings (PMH) algorithm \citep{AndrieuDoucetHolenstein2010}. The MH acceptance probability depends on the intractable likelihood, which in PMH is estimated using SMC (see Section~\ref{sec:overview}). Despite the apparent approximation, this results in an algorithm that targets the correct posterior distribution \citep{AndrieuDoucetHolenstein2010}. The original PMH algorithm makes use of a marginal proposal for $\theta$, i.e.\ only the current parameter is used when proposing a new parameter. The theoretical properties of the marginal PMH algorithm have been analysed in \citet{AndrieuVihola2012,PittSilvaGiordaniKohn2012,DoucetPittKohn2012} and it has been applied for a number of interesting applications in, e.g.,\ economics, social network analysis and ecology \citep{FluryShephard2011,Everitt2012,GolightlyWilkinson2011}.

In this paper, we show that information such as the gradient and the Hessian about the posterior can be included in the construction of the PMH proposal. This idea is first suggested by \citet{DoucetJacobJohansen2011} in the discussions following \citet{GirolamiCalderhead2011}. In two previous proceedings, we have applied and extended this idea with gradient information \citep{DahlinLindstenSchon2013a} and also using Hessian information \citep{DahlinLindstenSchon2014a}. The present article builds upon and extends this preliminary work. A PMH method using gradient information similar to \citet{DahlinLindstenSchon2013a} has recently been proposed by \citet{NemethFearnhead2014}.

In the context of MH sampling, it has been recognised that the gradient and Hessian can be used to construct efficient proposal distributions. In the Metropolis adjusted Langevin algorithm (MALA) \citep{RobertsStramer2003}, a drift term is added to the proposal in the direction of the gradient, which intuitively guides the Markov chain to regions of high posterior probability. In the manifold MALA (mMALA) \citep{GirolamiCalderhead2011}, the Hessian (or some other appropriate metric tensor) is also included to scale the proposal to take the curvature of the log-posterior into account. Drawing parallels with the optimisation literature, mMALA shares some properties with Newton-type optimisation algorithms (where MALA is more similar to a steepest ascent method). In particular, scaling the proposal with the Hessian can considerably simplify the tedious tuning of the method since it removes the need for running costly pilot runs, which are commonly used to tune the covariance matrices of the random walk MH and the MALA.

In our problem, i.e.\ for inference in a nonlinear SSM \eqref{eq:SSM}, the gradient and Hessian cannot be computed analytically. However, in analogue with the intractable likelihood, these quantities can be estimated using SMC algorithms, see e.g.\ \citet{PoyiadjisDoucetSingh2011,DoucetJacobRubenthaler2013}. This provides us with the tools necessary to construct PMH algorithms in the flavour of the MALA and the mMALA, resulting in the two methods proposed in this paper, PMH1 and PMH2, respectively. In particular, we make use of a fixed-lag (FL) particle smoother \citep{KitagawaSato2001} to estimate the gradient and Hessian. The motivation for this is that this smoother only makes use of the weighted particles computed by the particle filter. Consequently, we obtain this information as a \textit{byproduct} of the likelihood computation in the PMH algorithm. This results in only a small computational overhead for the proposed methods when compared to the marginal method.

Finally, we provide numerical experiments to illustrate the benefits of using the gradient and Hessian and the accuracy of the FL smoother. We demonstrate some interesting properties of the proposed algorithms, in particular that they enjoy (i) a shorter burn-in compared with the marginal algorithm, (ii) a better mixing of the Markov chain in the stationary phase, and (iii) a simplified tuning of the step length(s), especially when the target distribution is non-isotropic.


\section{Particle Metropolis-Hastings}
\label{sec:overview}
In this section, we review the PMH algorithm and show how the random variables used to compute the likelihood estimator can be incorporated in the proposal construction. We also outline the idea of how this can be used to construct the proposed PMH1 and PMH2 algorithms.

\subsection{MH sampling with unbiased likelihoods}
\label{sec:PMH}
The MH algorithm (see, e.g.\ \citet{RobertCasella2004}) is a member of the MCMC family for sampling from a target distribution $\pi(\theta)$ by simulating a carefully constructed Markov chain on $\Theta$. The chain is constructed in such a way that it admits the target as its unique stationary distribution. 

The algorithm consists of two steps: (i) a new parameter $\theta''$ is sampled from a proposal distribution $q(\theta''|\theta')$ given the current state $\theta'$ and (ii) the current parameter is changed to $\theta''$ with probability $\alpha(\theta',\theta'')$, otherwise the chain remains at the current state. The acceptance probability is given by
\begin{align}
	\alpha(\theta',\theta'') 
	= 
	1 \wedge
	\frac{ \pi(\theta'') }{\pi(\theta')} 
	\frac{ q(\theta' | \theta'')}{q(\theta'' | \theta')},
	\label{eq:MHacceptprob}
\end{align}
where we use the notation $a \wedge b \triangleq \min\{a,b\}$.

In this paper, we have the parameter posterior distribution \eqref{eq:parameterPosterior} as the target distribution, i.e.\ $\pi(\theta) = p(\theta|\mathbf{y})$. This implies that the acceptance probability \eqref{eq:MHacceptprob} will depend explicitly on the intractable likelihood $p_\theta(\mathbf{y})$, preventing direct application of the MH algorithm to this problem. However, this difficulty can be circumvented by using a \emph{pseudo-marginal} approach \citep{Beaumont2003,AndrieuRoberts2009}.

Assume that there exists an unbiased, non-negative estimator of the likelihood $\widehat{p}_{\theta}(\mathbf{y}|u)$. We introduce explicitly the random variable $u \in \mathsf{U}$ used to construct this estimator, and we let $m_{\theta}(u)$ denote the probability density of $u$ on $\mathsf{U}$.

The pseudo-marginal method is then a standard MH algorithm
operating in a non-standard extended space $\Theta \times \mathsf{U}$, with the \textit{extended target}
\begin{align*}
	\pi( \theta, u | \mathbf{y} ) 
	= 
	\frac{ \widehat{p}_{\theta}( \mathbf{y} | u ) m_{\theta}(u) p(\theta) } { p(\mathbf{y}) } 
	=
	\frac{ \widehat{p}_{\theta}( \mathbf{y} | u ) m_{\theta}(u) p(\theta | \mathbf{y}) } { p_{\theta}(\mathbf{y}) },
\end{align*}
and proposal distribution $m_{\theta''}(u'') q(\theta'' | \theta')$.

Since the likelihood estimator is unbiased, $\mathbb{E}_{u|\theta}[\widehat{p}_{\theta}(\mathbf{y}|u)] = p_{\theta}( \mathbf{y} )$, it follows that the extended target admits $p(\theta | \mathbf{y})$ as a marginal. Hence, by simulating from the extended target $\pi( \theta, u | \mathbf{y} ) $ we obtain samples from the original target distribution $p(\theta | \mathbf{y})$ as a byproduct.

If the likelihood is estimated by using SMC (see Section~\ref{sec:SMC}) we obtain the
PMH algorithm. The random variable $u$ then corresponds to all the weighted particles generated by the SMC algorithm. However, these random variables carry useful information, not only about the likelihood, but also about the geometry of the posterior distribution. We suggest to incorporate this information into the proposal construction. With $(\theta', u')$ being the current state of the Markov chain we simulate $\theta'' \sim q(\cdot | \theta', u')$ and $u'' \sim m_{\theta''}(\cdot)$, using some proposal $q$  (see Section~\ref{sec:constructPMH2}).

It follows that the (standard) MH acceptance probability for the extended target is given by
\begin{align}
  \nonumber
  \alpha(\theta'',u'', \theta', u')
  &=  1  \wedge
  \frac{ \widehat{p}_{\theta''}(\mathbf{y} | u'') m_{\theta''}(u'') p(\theta'')}{ \widehat{p}_{\theta'}(\mathbf{y} | u') m_{\theta'}(u') p(\theta') }
  \frac{ m_{\theta'}(u') q(\theta' | \theta'', u'') }{ m_{\theta''}(u'') q(\theta'' | \theta', u' )} \\
  &=1  \wedge
  \frac{ \widehat{p}_{\theta''}(\mathbf{y} | u'') p(\theta'')}{ \widehat{p}_{\theta'}(\mathbf{y} | u') p(\theta') }
  \frac{ q(\theta' | \theta'', u'') }{ q(\theta'' | \theta', u' )}.
  \label{eq:PMHaprob}
\end{align}
Note that $ q(\theta'' | \theta', u' )$ may depend on the auxiliary variable $u'$ in a (formally) arbitrary way.
In particular, in Section~\ref{sec:SMC} we propose a construction making use of \emph{biased} estimates of the
gradient and Hessian of the log-posterior. Nevertheless, expression \eqref{eq:PMHaprob} still defines
a correct MH acceptance probability for the extended target, ensuring the validity of our approach. Note also that the aforementioned proposal construction opens up for a wide range of adapted proposals, possibly different from the ones considered in this work.

\subsection{Constructing PMH1 and PMH2}
\label{sec:constructPMH2}
We now turn to the construction of a proposal that makes use of the gradient and Hessian of the log-posterior. Following \citet{RobertCasella2004}, we do this by a Laplace approximation of the parameter posterior around the current state $\theta'$. Hence, consider a second order Taylor expansion of $\log p(\theta''|\mathbf{y})$ at $\theta'$:
\begin{align*}
	\log p(\theta''|\mathbf{y})
	&\approx 	
	\log p(\theta'|\mathbf{y}) 
	+ (\theta'' - \theta')^{\top} \Big[ \Dtheta \log p(\theta|\mathbf{y}) \Big]_{\theta=\theta'} \\
	&+ \frac{1}{2} (\theta'' - \theta')^{\top} \Big[ \DDtheta \log p(\theta|\mathbf{y}) \Big]_{\theta=\theta'} (\theta'' - \theta').
\end{align*}
Taking the exponential of both sides and completing the square, we obtain
\begin{align*}
	p(\theta''| \mathbf{y})
	&\approx
	\textsf{N} \Big( \theta'';\theta' + \mathsf{I}^{-1}_T(\theta') \mathsf{S}_T(\theta'), \mathsf{I}^{-1}_T(\theta') \Big),
\end{align*}
where we have introduced $\mathsf{S}_T(\theta') = \Dtheta \log p(\theta|\mathbf{y})|_{\theta=\theta'}$ and $\mathsf{I}_T(\theta') = - \DDtheta \log p(\theta|\mathbf{y})|_{\theta=\theta'}$, for the gradient
and the negative Hessian of the log-posterior, respectively. Here, we assume for now that the negative Hessian is positive definite; see Section~\ref{sec:regularisation} for further discussion on this matter. 

As pointed out above, these quantities cannot be computed in closed form, but they can be estimated from the random variable $u'$ (see Section~\ref{sec:SMC}). This suggests three different versions of the PMH algorithm, each resulting from a specific choice of the proposal:
\begin{align}
q(\theta''|\theta',u') =
\begin{dcases}
\textsf{N}\left( \theta', \Gamma \right), & \text{[PMH0]}  \\
\textsf{N}\left( \theta' + {\textstyle \frac{1}{2}} \Gamma \widehat{\mathsf{S}}_T(\theta'|u'), \Gamma \right), & \text{[PMH1]} \\
\textsf{N}\left( \theta' + \widehat{\textsf{G}}(\theta'|u'), \widehat{\textsf{H}}(\theta'|u') \right). & \text{[PMH2]}
\end{dcases}
\label{eq:2orderproposal}
\end{align}
Here, we use the notation $\widehat{\textsf{G}}(\theta|u) = \frac{1}{2} \Gamma\, \widehat{\mathsf{I}}^{-1}_T(\theta|u) \, \widehat{\mathsf{S}}_T(\theta|u)$ and $\widehat{\textsf{H}}(\theta|u) = \Gamma\, \widehat{\mathsf{I}}^{-1}_T(\theta|u)$ for the natural gradient and scaled inverse Hessian, respectively. Furthermore, $\Gamma$ denotes a scaling matrix that controls the step lengths of the proposal. For PMH0 and PMH1, $\Gamma$ can be chosen as the inverse of an estimate of the posterior covariance matrix. However, computing this estimate typically requires costly and tedious trial runs. For PMH2, the curvature of the problem is captured by the Hessian matrix, i.e.\ a single step length can by used which can significantly simplify the tuning. It is also possible to choose different step lengths for the drift term and for the covariance matrix of the proposal.

The final PMH2 algorithm is presented in Algorithm~\ref{alg:PMH2order}. It makes use of Algorithm~\ref{alg:SMCfull}, described in Section~\ref{sec:SMC}, to estimate the quantities needed for computing the proposal and the acceptance probability. Clearly, PMH0 and PMH1 are special cases obtained by using the corresponding proposal from \eqref{eq:2orderproposal} in the algorithm. Note that, while the algorithm make explicit reference to the auxiliary variable $u$, it only depends on this variable through the estimates $\widehat{p}_{\theta'}(\mathbf{y})$, $\widehat{\mathsf{S}}_T(\theta')$ and $\widehat{\mathsf{I}}_T(\theta')$.

\begin{algorithm}[!t]
\caption{\textsf{Second order particle Metropolis-Hastings}}
\textsc{Inputs:} Algorithm~\ref{alg:SMCfull}. $M>0$ (no.\ MCMC steps), $\theta_0$ (initial parameters), $\gamma$ (step length). \\
\textsc{Output:} $\theta=\{\theta_1,\ldots,\theta_M\}$ (samples from the posterior).
\algrule[.4pt]
\begin{algorithmic}[1]
	\STATE Run Algorithm \ref{alg:SMCfull} to obtain $\widehat{p}_{\theta_0}(\mathbf{y})$, $\widehat{\mathsf{S}}_T(\theta_0)$ and $\widehat{\mathsf{I}}_T(\theta_0)$.
	\FOR{$k=1$ to $M$}
		\STATE Sample $\theta' \sim q(\theta'|\theta_{k-1},u_{k-1})$ by \eqref{eq:2orderproposal},  $\widehat{\mathsf{S}}_T(\theta_{k-1})$ and $\widehat{\mathsf{I}}_T(\theta_{k-1})$.
		\STATE Run Algorithm \ref{alg:SMCfull} to obtain $\widehat{p}_{\theta'}(\mathbf{y})$, $\widehat{\mathsf{S}}_T(\theta')$ and $\widehat{\mathsf{I}}_T(\theta')$.
		\STATE Sample $\omega_k$ uniformly over $[0,1]$.
		\IF{$\omega_k < \alpha(\theta',u',\theta_{k-1},u_{k-1})$ given by \eqref{eq:PMHaprob}}
			\STATE $\theta_k \leftarrow \theta'$. \COMMENT{Accept the parameter}
			\STATE $\{ \widehat{p}_{\theta_k}(\mathbf{y}), \widehat{\mathsf{S}}_T(\theta_{k}), \widehat{\mathsf{I}}_T(\theta_{k}) \} \leftarrow \{ \widehat{p}_{\theta'}(\mathbf{y}) , \widehat{\mathsf{S}}_T(\theta') , \widehat{\mathsf{I}}_T(\theta') \}$.
		\ELSE
			\STATE $\theta_k \leftarrow \theta_{k-1}$. \COMMENT{Reject the parameter}
			\STATE $\{ \widehat{p}_{\theta_k}(\mathbf{y}), \widehat{\mathsf{S}}_T(\theta_{k}), \widehat{\mathsf{I}}_T(\theta_{k}) \} \leftarrow \{ \widehat{p}_{\theta_{k-1}}(\mathbf{y}) , \widehat{\mathsf{S}}_T(\theta_{k-1}) , \widehat{\mathsf{I}}_T(\theta_{k-1}) \}$.
		\ENDIF
	\ENDFOR
\end{algorithmic}
\label{alg:PMH2order}
\end{algorithm}

\subsection{Properties of the PMH1 and PMH2 proposals}
In the sequel, we use a single step size $\Gamma = \gamma^2 I_d$ for all the parameters in the (standard) proposal. This is done to illustrate the advantage of adding the Hessian information, which rescales the step lengths according to the local curvature. Hence, it allows for taking larger steps when the curvature is small and vice verse. 

This property of PMH2 makes the algorithm scale-free in the same manner as a Newton algorithm in optimisation \citep[Chapter 3]{NocedalWright2006}. That is, the proposal is invariant to affine transformations of the parameters. Note that, since the local information is used, this is different from scaling the proposal in PMH0 with the posterior covariance matrix estimated from a pilot run, as this only takes the geometry at the mode of the posterior into account.

Some analyses of the statistical properties are available for PMH0 \citep{SherlockThieryRobetsRosenthal2013}, MH using a random walk \citep{RobertsGelmanGilks1997} and MALA \citep{RobertsRosenthal1998}. It is known from these analyses that adding the gradient into the proposal can increase the mixing of the Markov chain. Note that these results are obtained under somewhat strict assumptions. Also, we know from numerical experiments \citep{GirolamiCalderhead2011} that there are further benefits of also taking the local curvature into account.


\section{Estimation of the likelihood, gradient, and Hessian}
\label{sec:SMC}
In this section, we show how to estimate the likelihood together with the gradient and Hessian using SMC methods.

\subsection{Auxiliary particle filter}
An auxiliary particle filter (APF) \citep{PittShephard1999} can be used to approximate the sequence of joint smoothing distributions (JSDs) $p_{\theta}(x_{1:t}|y_{1:t})$ for $t = 1$ to $T$. The APF makes use of a particle system consisting of $N$ weighted particles $\{x_{1:t}\pIdx{i},w_t\pIdx{i}\}_{i=1}^N$ to approximate the JSD at time $t$ by
\begin{align}
	\widehat{p}_{\theta}(\dd x_{1:t} | y_{1:t} ) 
	\triangleq 
	\sum_{i=1}^N 
	\frac 
	{ w_t\pIdx{i}  }
	{ \sum_{k=1}^N w_t\pIdx{k} }
	\delta_{x_{1:t}\pIdx{i}} (\dn x_{1:t}).
	\label{eq:empericalfiltering}
\end{align}
Here, $\delta_z(\dn x_{1:t})$ denotes the Dirac measure placed at $z$. The particle system is propagated from $t-1$ to $t$ by first sampling an \emph{ancestor index} $a_t\pIdx{i}$, with
\begin{align}
 \mathbb{P}( a_t\pIdx{i} = j ) = \nu_{t-1}^{(j)} \left[ \sum_{k=1}^N \nu\pIdx{k}_{t-1} \right]^{-1},
\quad i,j = 1,\dots,N,
	\label{eq:APFresampling} 
\end{align}
where $\nu\pIdx{i}_{t-1}$ denotes the resampling weights. Given the ancestor index, a new particle is sampled according to
\begin{align}
	\quad
	x_t\pIdx{i} \sim R_{\theta} \Big( x_t | x^{a_t\pIdx{i}}_{1:t-1}, y_t \Big),
	\quad i = 1,\dots,N.
	\label{eq:APFpropagation} 
\end{align}
Finally, we append the obtained sample to the trajectory by $x_{1:t}\pIdx{i}=\{x_{1:t-1}^{a_t\pIdx{i}},x_t\pIdx{i}\}$ and compute a new importance weight by
\begin{align}
	w_{t}\pIdx{i}
	&\triangleq 
	\frac
	{w_{t-1}^{a\pIdx{i}_t}}
	{\nu_{t-1}^{a\pIdx{i}_t}}
	\frac
	{ g_{\theta} \Big( y_t \Big| x_t\pIdx{i} \Big) f_{\theta} \Big( x_t\pIdx{i} \Big| x_{t-1}^{a\pIdx{i}_t} \Big) }
	{ R_{\theta} \Big( x_t\pIdx{i} \Big| x_{1:t-1}^{a\pIdx{i}_t},y_t \Big) },
	\quad i = 1,\dots,N.
	\label{eq:APFweights}
\end{align}%
Hence, the empirical approximations of the smoothing distributions \eqref{eq:empericalfiltering} can be computed sequentially for $t=1$ to $T$ by repeating \eqref{eq:APFresampling}--\eqref{eq:APFweights}. 

Note that the random variables $u$ appearing in the extended target of the PMH algorithm correspond to all the random variables generated by the APF, i.e.\ all the particles and ancestor indices,
\begin{align*}
  u= \bigg( \Big\{x\pIdx{i}_{t}, a\pIdx{i}_{t} \Big\}_{i=1}^N, t = 1,\,\dots,\,T \bigg).
\end{align*}
In this article, we make use of two important special cases of the APF: the bootstrap particle filter (bPF) \citep{GordonSalmondSmith1993} and the fully adapted particle filter (faPF) \citep{PittShephard1999}. For the bPF, we select the proposal kernel  $R_{\theta}(x_t|x_{1:t-1},y_t) = f_{\theta}(x_t|x_{t-1})$ and the auxiliary weights $\nu_t = w_t = g_{\theta}(y_t|x_t)$. The faPF is obtained by $R_{\theta}(x_t|x_{1:t-1},y_t) = p_{\theta}(x_t|y_t,x_{t-1})$ and $\nu_t = p_{\theta}(y_{t+1}|x_t)$, resulting in the weights $w_t \equiv 1$. Note, that the faPF can only be used in models for which these quantities are available in closed-form.

\subsection{Estimation of the likelihood}
The likelihood for the SSM in \eqref{eq:SSM} can be estimated using \eqref{eq:likeFunc} by inserting estimated one-step predictors $p_{\theta}(y_t|y_{1:t-1})$ obtained from the APF. The resulting likelihood estimator is given by
\begin{align}
	\widehat{p}_{\theta}( \mathbf{y} | u ) 
	= \frac{1}{N^T} \sum_{i=1}^N w_{T}\pIdx{i}  \left\{ \prod_{t=1}^{T-1}
        \sum_{i=1}^N \nu_{t}\pIdx{i} \right\}.
	\label{eq:EstLikelihood}
\end{align}
\noindent It is known that this likelihood estimator is unbiased for any number of particles, see e.g.\ \citep{PittSilvaGiordaniKohn2012} and Proposition 7.4.1 in \citep{DelMoral2004}. As discussed in Section~\ref{sec:PMH}, this is exactly the property that is needed in order to obtain $p(\theta | \mathbf{y})$ as the unique stationary distribution for the Markov chain generated by the PMH algorithm.

Consequently, PMH will target the correct distribution for any number of particles $N\geq 1$. However, the variance in the likelihood estimate is connected with the acceptance rate and the mixing of the Markov chain. Therefore it is important to determine the number of particles that balances a reasonable acceptance rate with a reasonable computational cost. This problem is studied for PMH0 in \citet{PittSilvaGiordaniKohn2012,DoucetPittKohn2012}.

\subsection{Estimation of the gradient}
As we shall see below, the gradient of the log-posterior can be estimated by solving a smoothing problem. The APF can be used directly to address this problem, since the particles $\{x_{1:T}^{(i)}, w_T^{(i)}\}_{i=1}^N$ provide an approximation of the JSD at time $T$ according to \eqref{eq:empericalfiltering} (see also \citet{PoyiadjisDoucetSingh2011}). However, this method can give estimates with high variance due to \textit{particle degeneracy}. 

Instead, we make use of the FL smoother \citep{KitagawaSato2001} which has the same linear computational cost, but smaller problems with \textit{particle degeneracy} than the APF. Alternative algorithms for estimating this information are also available \citep{DelMoralDoucetSingh2010,PoyiadjisDoucetSingh2011}.

The gradient of the parameter log-posterior is given by
\begin{align}
	\mathsf{S}_T(\theta) = \Dtheta \log p(\theta) + \Dtheta \log p_{\theta}(\mathbf{y}),
	\label{eq:deffirst order}
\end{align}
where it is assumed that the gradient of the log-prior $\Dtheta \log p(\theta)$ can be calculated explicitly. The gradient of the log-likelihood $\Dtheta \log p_{\theta}(\mathbf{y})$ can, using \textit{Fisher's identity} \citep{CappeMoulinesRyden2005}, be expressed as
\begin{align}
	\Dtheta \log p_{\theta}(\mathbf{y}) 
	&= 
	\mathbb{E}_{\theta} \left[ \Dtheta \log p_{\theta}(\mathbf{x},\mathbf{y}) \Big| \mathbf{y} \right],
	\label{eq:FishersIdentity}	
\end{align}
where for an SSM \eqref{eq:SSM} we can write the gradient of the complete data log-likelihood as
\begin{align}
	\Dtheta \log p_{\theta}(\mathbf{x},\mathbf{y})
	&=
	\sum_{t=1}^T
	\xi_{\theta}(x_t,x_{t-1}), \text{ where}
	\label{eq:jointdistSSM} \\
	\xi_{\theta}(x_t,x_{t-1}) &= 
	\Dtheta \log f_{\theta}(x_t|x_{t-1}) + \Dtheta \log g_{\theta}(y_t|x_{t}). \nonumber
\end{align}
Combining \eqref{eq:jointdistSSM} with Fisher's identity \eqref{eq:FishersIdentity} yields
\begin{align*}
\Dtheta \log p_{\theta}(\mathbf{y}) 
&=
\sum_{t=1}^{T} \dint
\xi_{\theta}(x_{t}, x_{t-1})
p_{\theta}(x_{t-1:t}|\mathbf{y}) \dd x_{t-1:t},
\end{align*}
which depends on the (intractable) two-step smoothing distribution  $p_{\theta}(x_{t-1:t}|\mathbf{y})$. To approximate this quantity we use the FL smoother which relies on the assumption that there is a decaying influence of future observations $y_{t+\Delta:T}$ on the state $x_t$. This means that
\begin{align*}
p_{\theta}(x_{t-1:t}|\mathbf{y}) \approx p_{\theta}(x_{t-1:t}|y_{1:\kappa_t}),
\end{align*}
holds for some large enough $\kappa_t=\min\{t+\Delta,T\}$. Here, $\Delta$ denotes a pre-determined lag decided by the user, which depends on the forgetting properties of the model. By marginalisation of the empirical smoothing distribution $\widehat{p}_{\theta}(x_{1:\kappa_t}|y_{1:\kappa_t})$ over $x_{1:t-2}$ and $x_{t+1:\kappa_t}$, we obtain the approximation
\begin{align}
	\widehat{p}_{\theta}^\Delta(\dn x_{t-1:t}| \mathbf{y}) 
	\triangleq 
	\sum_{i=1}^N 
	w_{\kappa_t}\pIdx{i} 
	\delta_{\tilde{x}_{\kappa_t,t-1:t}\pIdx{i}}
	(\dn x_{t-1:t}).
	\label{eq:FL2step}
\end{align}
Here, we use the notation $\tilde{x}_{\kappa_t,t}^{(i)}$ to denote the ancestor at time~$t$ of particle $x_{\kappa_t}^{(i)}$ and $\tilde{x}_{\kappa_t,t-1:t}^{(i)} = \{ \tilde{x}_{\kappa_t,t-1}^{(i)}, \tilde{x}_{\kappa_t,t}^{(i)} \}$. Inserting \eqref{eq:jointdistSSM}--\eqref{eq:FL2step} into \eqref{eq:FishersIdentity} provides an estimator of \eqref{eq:deffirst order},
\begin{align}
	\widehat{\mathsf{S}}_T(\theta|u)
	&=
	\nabla \log p(\theta) +
	\sum_{t=1}^{T}
	\sum_{i=1}^N	
	w_{\kappa_t}\pIdx{i}
	\xi_{\theta} \Big( \tilde{x}_{\kappa_t,t}\pIdx{i}, \tilde{x}_{\kappa_t,t-1}\pIdx{i} \Big),
	\label{eq:FisherScoreParticleApproximation}
\end{align}
which is used in the proposal distributions in \eqref{eq:2orderproposal}.

\subsection{Estimation of the Hessian}
The negative Hessian of the parameter log-posterior can be written as
\begin{align}
	\mathsf{I}_T(\theta) = -\DDtheta \log p(\theta) - \DDtheta \log p_{\theta}( \mathbf{y} ),
		\label{eq:defSecondOrder}
\end{align}
where it is assumed that the Hessian of the log-prior $\DDtheta \log p(\theta)$ can be calculated analytically. The negative Hessian of the log-likelihood, also known as the \textit{observed information matrix}, can using \textit{Louis' identity} \citep{CappeMoulinesRyden2005} be expressed as
\begin{align}
	- \DDtheta \log p_{\theta}(\mathbf{y}) 
	&= 
	\Dtheta \log p_{\theta}(\mathbf{y})^2 
	-
    \mathbb{E}_{\theta} 
    \Big[ \DDtheta \log p_{\theta}(\mathbf{x},\mathbf{y}) \Big| \mathbf{y} \Big] 	\nonumber \\
    &-\mathbb{E}_{\theta} \Big[ 
    \Dtheta \log p_{\theta}(\mathbf{x},\mathbf{y})^2
    \Big| \mathbf{y} \Big].
    \label{eq:LouisIdentity}
\end{align}
Here, we have introduced the notation $v^2=vv^{\top}$ for a vector~$v$. From this, we can construct an estimator of \eqref{eq:defSecondOrder} using the estimator of the gradient in \eqref{eq:FisherScoreParticleApproximation}, of the form
\begin{align}
	\widehat{\mathsf{I}}_T(\theta|u) 
	=
	-\DDtheta \log p(\theta)
	+ \widehat{\mathsf{S}}_T(\theta|u)^2
	- \widehat{\mathsf{I}}^{(1)}_T(\theta|u)
	- \widehat{\mathsf{I}}^{(2)}_T(\theta|u),
	\label{eq:LouisIdentityEst}
\end{align}
where we introduce $\mathsf{I}^{(1)}_T(\theta)= \mathbb{E}_{\theta} \left[ \DDtheta \log p_{\theta}(\mathbf{x},\mathbf{y}) | \mathbf{y} \right]$ and \, ${\mathsf{I}^{(2)}_T(\theta)= \mathbb{E}_{\theta} \left[ \Dtheta \log p_{\theta}(\mathbf{x},\mathbf{y})^2| \mathbf{y} \right]}$. We obtain the estimator of the first term
analogously to \eqref{eq:FisherScoreParticleApproximation} as
\begin{align}
	\widehat{\mathsf{I}}^{(1)}_{T}(\theta|u)
	&= \sum_{t=1}^T \sum_{i=1}^N 
	w_{\kappa_t}\pIdx{i}  \zeta_{\theta} \Big( \tilde{x}_{\kappa_t,t}\pIdx{i}, \tilde{x}_{\kappa_t,t-1}\pIdx{i} \Big), \text{ where} \label{eq:LouisIdentityEstTerm1Final} \\
	\zeta_{\theta}(x_t, x_{t-1})
	&=
	\DDtheta \log f_{\theta}(x_t|x_{t-1}) + \DDtheta \log g_{\theta}(y_t|x_{t}). \nonumber
\end{align}
The estimator of the second term needs a bit more work and we start by rewriting the last term in \eqref{eq:LouisIdentity} as
\begin{align}
&\sum_{t=1}^T \sum_{s=1}^T \mathbb{E}_{\theta} \left[ \xi_{\theta}(x_t,x_{t-1}) \xi_{\theta}(x_s,x_{s-1})^{\top} \Big| \mathbf{y} \right] \nonumber \\
&=
\sum_{t=1}^T 
\bigg\{ 
\mathbb{E}_{\theta} \left[ \xi_{\theta}(x_t,x_{t-1})^2 \Big| \mathbf{y} \right] \nonumber \\
&+ \sum_{s=1}^{t-1} 
\mathbb{E}_{\theta} \left[ \big( \xi_{\theta}(x_t,x_{t-1}),\xi_{\theta}(x_s,x_{s-1}) \big)^{\dagger} \Big| \mathbf{y} \right]
\bigg\}, \label{eq:LouisIdentityEstTerm2}
\end{align}
where we have introduced the operator $(a,b)^{\dagger}=ab^{\top}+ba^{\top}$ for brevity.
Consider the last term appearing in this expression, we can rewrite it as
\begin{align*}
&\sum_{s=1}^{t-1} 
\mathbb{E}_{\theta} \left[ \xi_{\theta}(x_t,x_{t-1}) \xi_{\theta}(x_s,x_{s-1})^{\top} \Big| \mathbf{y} \right] \\
&=
\mathbb{E}_{\theta} \Bigg[ \xi_{\theta}(x_t,x_{t-1}) 
\underbrace{\left\{ \sum_{s=1}^{t-1} 
\mathbb{E}_{\theta} \left[ \xi_{\theta}(x_s,x_{s-1}) 
\big| x_{t-1}, y_{1:t-1} \right] \right\}^{\top}}
_{\triangleq \alpha_{\theta}(x_{t-1})^{\top}}
 \Big| \mathbf{y} \Bigg].
\end{align*}
From this, we see that \eqref{eq:LouisIdentityEstTerm2} can be written as an additive functional of the form
\begin{align*}
\sum_{t=1}^T 
\mathbb{E}_{\theta} \left[ 
(\xi_{\theta}(x_t,x_{t-1}))^2 
+ 
\big(( \xi_{\theta}(x_t,x_{t-1}),\alpha_{\theta}(x_{t-1}) \big)^{\dagger} 
\Big| \mathbf{y} 
\right],
\end{align*}
which can be estimated using the FL smoother as before. However, for this we need to compute the quantities $\alpha_{\theta}(x_{t-1})$. One option is to make use of a type of fixed-lag approximation for $\alpha_{\theta}(x_{t-1})$, by assuming that $x_s$ and $x_t$ are conditionally independent given $y_{1:\kappa_t}$, whenever $|s-t| > \Delta$. This approach has previously been used by \citet{DoucetJacobRubenthaler2013}. Alternatively, we can use a filter approximation according to
\begin{align}
	\widehat{\alpha}_{\theta} \Big( x_{t}\pIdx{i} \Big) 
	= 
	\widehat{\alpha}_{\theta} \Big( x_{t-1}^{a_{t}\pIdx{i}} \Big) + 
	\xi_{\theta} \Big( x_t\pIdx{i},x_{t-1}^{a_{t}\pIdx{i}} \Big),
	\label{eq:alphaEst}
\end{align}
for $i=1, \ldots, N$. Note that this approach suffers from the same particle degeneracy as the APF. However, this only affects a small number of terms and in our experience this approximation works sufficiently well to give estimates with reasonable variance. The resulting estimator using \eqref{eq:LouisIdentityEstTerm2} is
\begin{align}
	\widehat{\mathsf{I}}^{(2)}_{T}(\theta|u)
	&= \sum_{t=1}^T \sum_{i=1}^N 
	w_{\kappa_t}\pIdx{i} \eta_{\theta} \Big( \tilde{x}_{\kappa_t,t}\pIdx{i}, \tilde{x}_{\kappa_t,t-1}\pIdx{i} \Big), \label{eq:LouisIdentityEstTerm2Final} \text{ where} \\
\eta_{\theta}(x_t, x_{t-1})
&=
\xi_{\theta}(x_t, x_{t-1})^2 + \big( \xi_{\theta}(x_t, x_{t-1}),\widehat{\alpha}_{\theta}(x_{t-1}) \big)^{\dagger}. \nonumber
\end{align}
Hence, the Hessian can be estimated using \eqref{eq:LouisIdentityEst} by inserting the estimators from \eqref{eq:LouisIdentityEstTerm1Final}, \eqref{eq:alphaEst} and \eqref{eq:LouisIdentityEstTerm2Final}.

\subsection{Regularisation of the estimate of the Hessian}
\label{sec:regularisation}
The PMH2 proposal \eqref{eq:2orderproposal} relies on the assumption that the observed information matrix is positive definite (PD). The estimator given in \eqref{eq:LouisIdentityEst} does not always satisfy this, especially when the Markov chain is located far from the posterior mode. Typically, the amount of information is limited in such regions and this results in that the curvature is difficult to estimate. To cope with this issue, one alternative is to regularize the Hessian by adding a diagonal matrix to shift the eigenvalues to be positive. The diagonal matrix can e.g.\ be selected such that
\begin{align}
	\Delta \widehat{I}_T = \max \Big\{0, - 2 \lambda_{\min} \big( \widehat{I}_T \big) \Big\} I_d,
	\label{eq:Hessianregularization}
\end{align}
where $\lambda_{\min}(\widehat{I}_T)$ denotes the smallest eigenvalue of $\widehat{I}_T(\theta|u)$. In this article, we make use of this method for handling non--PD estimates of the negative Hessian for the PMH2 algorithm. This heuristic is common for Newton-type optimisation algorithms \citep[Chapter 3.4]{NocedalWright2006}. 

Note, that there are other solutions available for ensuring positive definiteness that only shifts the negative eigenvalues, see \cite[Chapter 3]{NocedalWright2006}. We emphasise that this type of regularization keeps the Markov chain invariant, i.e.\ still targets the correct posterior distribution (recall Section~\ref{sec:PMH}).

Another alternative is to replace the estimate of the negative Hessian with the inverse sample covariance matrix calculated using the trace of Markov chain when the estimate is not PD. This can be seen as a hybrid between the PMH2 algorithm and a \textit{pre-conditioned PMH1 algorithm}. This resembles some other adaptive MH algorithms \citep{AndrieuThoms2008} in which the same procedure is used to adapt the covariance matrix of a random walk proposal. For this, we can make use of the last $L$ iterations of the MH algorithm after that the Markov chain has reached stationarity. During the burn-in phase, non--PD estimates can be handled using a regularization approach or by rejecting the proposed parameter. In this article, we refer to this method for handling non--PD estimates of the negative Hessian as the \textit{hybrid PMH2 algorithm}, where we use the latter alternative during the burn-in phase. Note that this pre-conditioning can also be applied to the PMH0 and PMH1 algorithm, we return to this in Section~\ref{sec:results:earth}.

\begin{algorithm}[!t]
\caption{\textsf{Estimation of the likelihood, the gradient and the Hessian of the log-posterior}}
\textsc{Inputs:} $\mathbf{y}$ (data), $R(\cdot)$ (propagation kernel), $\nu(\cdot)$ (weight function), $N > 0$ (no.\ particles), $0 < \Delta \leq T$ (lag). \\
\textsc{Outputs:} $\widehat{p}_{\theta}(\mathbf {y})$ (est.\ of the likelihood), $\widehat{\mathsf{S}}_T(\theta)$ (est.\ of the gradient), $\widehat{\mathsf{I}}_T(\theta)$ (est.\ of the negative Hessian).
\algrule[.4pt]
\begin{algorithmic}[1]
	\STATE Initialise each particle $x_0\pIdx{i}$.
	\FOR{$t=1$ to $T$}
		\STATE Resample and propagate each particle using \eqref{eq:APFpropagation}.
		\STATE Calculate the weights for each particle using \eqref{eq:APFweights}.
	\ENDFOR
	\STATE Compute $\widehat{p}_{\theta}(\mathbf{y})$ by \eqref{eq:EstLikelihood}.
	\STATE Compute $\widehat{\mathsf{S}}_T(\theta)$ and $\widehat{\mathsf{I}}_T(\theta)$ by \eqref{eq:FisherScoreParticleApproximation} and \eqref{eq:LouisIdentityEst}, respectively.
	\IF{$\widehat{\mathsf{I}}_T(\theta) \leq 0$}
		\STATE \textsf{[standard]} Regularize $\widehat{\mathsf{I}}_T(\theta)$ by adding $\Delta\widehat{\mathsf{I}}_T$ computed by \eqref{eq:Hessianregularization} 
		\STATE \textsf{[hybrid]} Replace $\widehat{\mathsf{I}}_T(\theta)$ by the inverse covariance matrix computed using the $L$ final samples of the Markov chain during the burn-in.
	\ENDIF
\end{algorithmic}
\label{alg:SMCfull}
\end{algorithm}

\subsection{Resulting SMC algorithm}
In Algorithm \ref{alg:SMCfull}, we present the complete procedure that combines the APF with the FL smoother to compute the estimates needed for the PMH2 proposal \eqref{eq:2orderproposal}. Note that the two different methods to handle non--PD estimates of the negative Hessian matrix results in the \textit{standard} and \textit{hybrid }PMH2 algorithm, respectively.

We end this section by briefly discussing the statistical properties of the estimates of the gradient and Hessian obtained from the FL smoother. From \citet{OlssonCappeDoucMoulines2008}, we know that the FL smoother gives biased estimates of the gradient and Hessian for any number of particles. Remember that this does not effect the invariance of the Markov chain (recall Section~\ref{sec:PMH}). The main advantage of the FL smoother over the APF (which gives a consistent estimate) is that the former enjoys a smaller variance than the APF, i.e.\ we obtain a favourable bias-variance trade-off for a certain choice of lag $\Delta$. Note that a too small lag gives a large bias in the estimate and a too large lag gives a large variance in the estimate; we return to this choice in Section~\ref{sec:results}.

\section{Numerical illustrations} \label{sec:results}
In this section, we provide illustrations of the properties of the FL smoother and the different proposed algorithms. The source code in Python and the data used for some of the numerical illustrations are available for download at: \url{http://liu.johandahlin.com/}.

\subsection{Estimation of the log-likelihood and the gradient}
\label{sec:results:flsmoother}
We begin by illustrating the use of the FL smoother for estimating the log-likelihood and the gradient. Here, we consider a linear Gaussian state space (LGSS) model given by
\begin{subequations}
\begin{align}
	x_{t+1}|x_{t} &\sim \textsf{N} \Big( x_{t+1}; \phi x_{t}, \sigma_v^2 \Big), \\
	y_{t}  |x_{t} &\sim \textsf{N} \Big( y_{t};   x_{t}, \sigma_e^2 \Big).
\end{align}
\label{eq:lgss}
\end{subequations}
We generate two data realisations of length $T=100$ using parameters $\theta^{(1)} = \{\phi,\sigma_v^2,\sigma_e^2\} = \{0.5,1.0,0.1^2\}$ and $\theta^{(2)} = \{0.5,1.0,1.0\}$ with a known initial zero state. We use the lag $\Delta = 5$ and run the PFs with systematic resampling \citep{CarpenterCliffordFearnhead1999}.

\begin{figure}[p]
	\centering
	\includegraphics[width=\columnwidth]{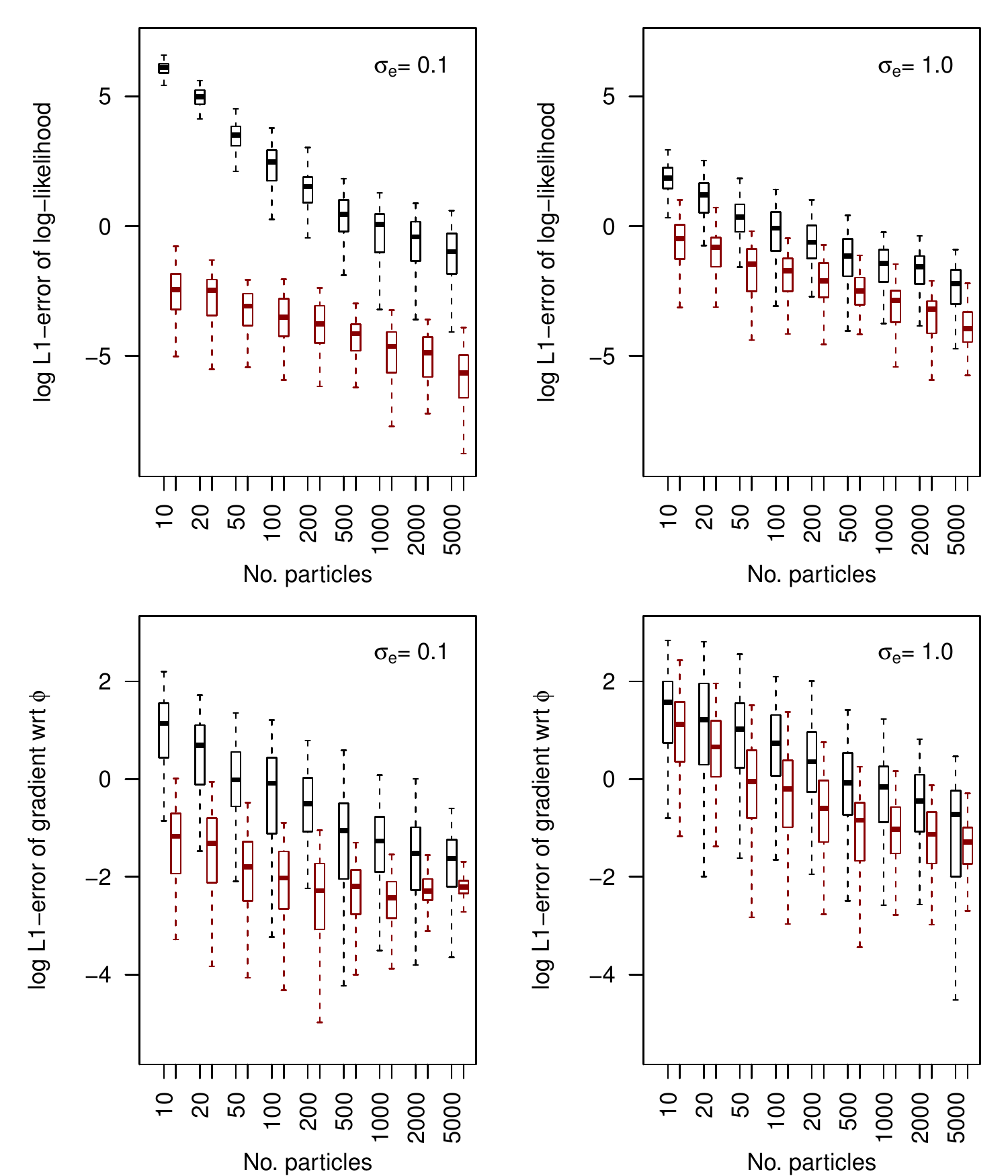}
	\caption{The log $L_1$-error in the log-likelihood estimates and the estimates of the gradient with respect to $\phi$ in the LGSS model with $\sigma_e=0.1$ (left) and $\sigma_e=1$ (right). The bPF (black) and faPF (red) are evaluated by $1 \thinspace 000$ MC iterations using a fixed data set with $T=100$.}
	\label{fig:score-lgss-n-paper}
\end{figure}

For this model, we can compute the true values of the log-likelihood and the gradient by running an RTS smoother \citep{RauchTungStriebel1965}. In Figure \ref{fig:score-lgss-n-paper}, we present boxplots of the $L_1$-errors in the estimated log-likelihood and the gradient of the log-posterior with respect to $\phi$, evaluated at the true parameters. When $\sigma_e=0.1$, we observe that the faPF has a large advantage over the bPF for all choices of $N$. When $\sigma_e=1.0$, we get smaller difference in the error of the gradient estimates, but the log-likelihood estimates are still better for the faPF. Similar results are also obtained for the gradient with respect to $\sigma_v$.  

\begin{figure}[p]
	\centering
	\includegraphics[width=\columnwidth]{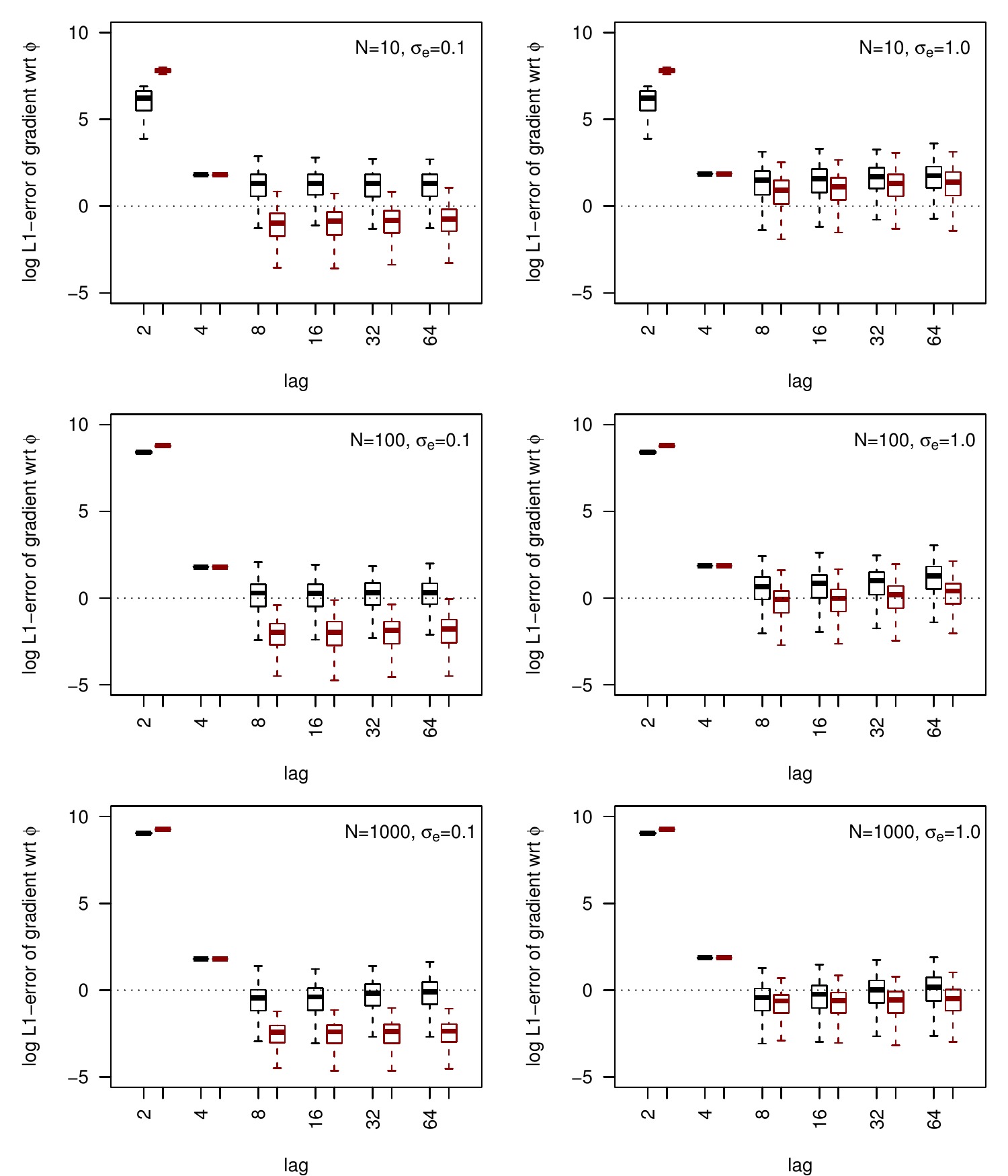}
	\caption{The log $L_1$-error in the estimates of the gradient with respect to $\phi$ in the LGSS model with $\sigma_e=0.1$ (left) and $\sigma_e=1$ (right). The bPF (black) and faPF (red) are evaluated by $1 \thinspace 000$ Monte Carlo iterations using a fixed data set with $T=100$.}
	\label{fig:score-lgss-lag-paper}
\end{figure}

In Figure \ref{fig:score-lgss-lag-paper}, we present the error in the gradient estimates with
respect to $\phi$ using a varying lag $\Delta$ and a varying number of particles $N$. The results are obtained by $1 \thinspace 000$ Monte Carlo runs on a single data set generated from the previously discussed LGSS model with $T=100$. We conclude again that faPF is preferable when available. The results are largely robust to the lag, as long as this is chosen large enough when using the faPF. A lag of about $12$ seems to be a good choice for this model when $T=100$ and when using the faPF with systematic resampling. 

\subsection{Burn-in and scale-invariance}
\begin{figure}[p]
	\centering
	\includegraphics[width=\columnwidth]{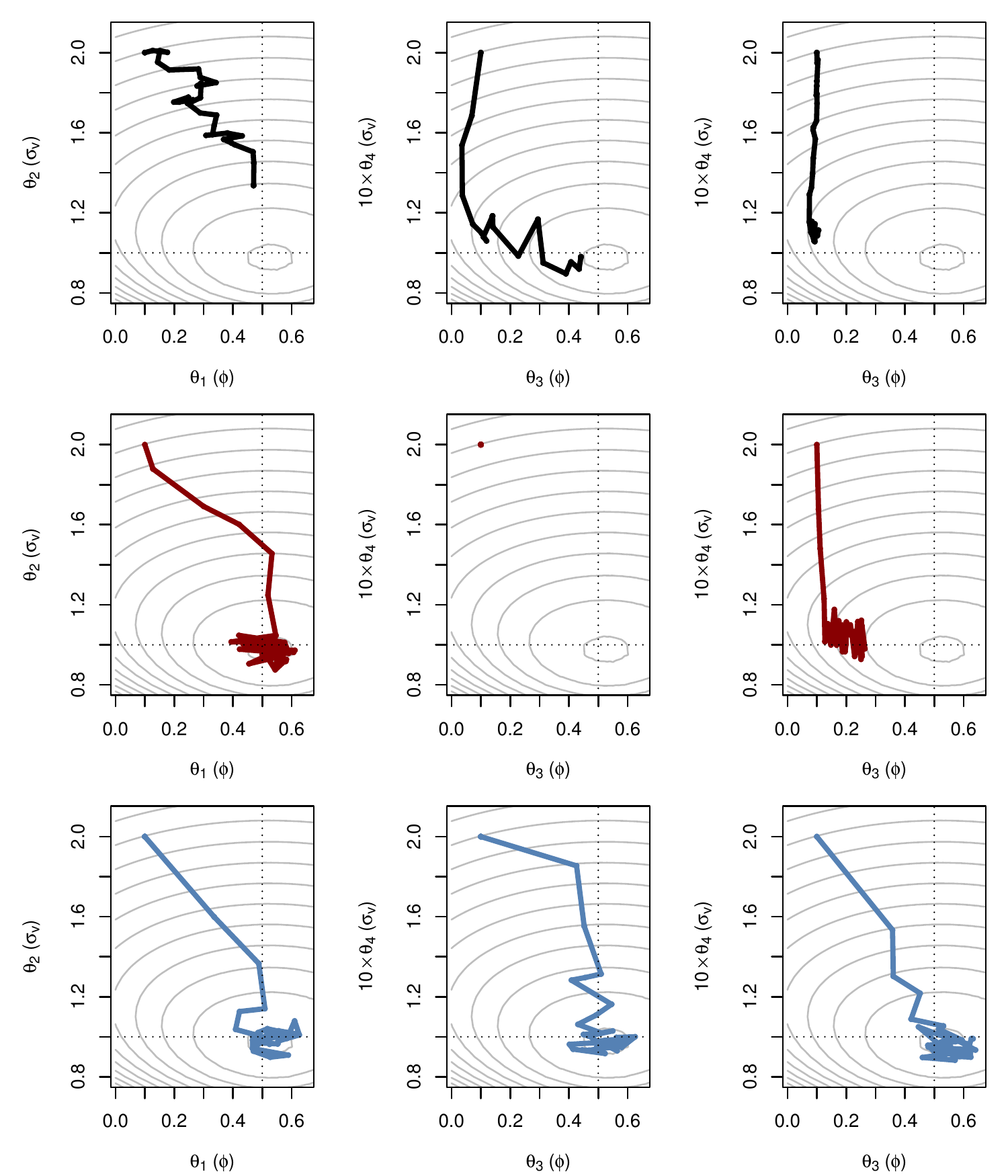}
	\caption{The trace plots of the first $50$ steps using PMH0 (black), PMH1 (red) and PMH2 (blue). The dotted lines show the \textit{true} parameters of the LGSS model. The gray contours show the log-posterior.}
	\label{fig:lgss-scaleinvariance}
\end{figure}

Consider the problem of inferring $\{\theta_1,\theta_2\}=\{\phi,\sigma_v\}$ in the LGSS model \eqref{eq:lgss}. We simulate a single data set with parameters $\theta^{(1)}$ (as defined in the previous section) of length $T=250$. We use an uniform parameter prior over $|\phi|<1,\sigma_v > 0$ and initialise in $\theta_0=\{0.1,2\}$. We use faPF with systematic resampling, $N=100$ and $\Delta=12$. Here, we use the standard version of Algorithm~\ref{alg:SMCfull} to adjust the estimate of the Hessian in the cases when it is not PD, resulting in the PMH2 algorithm.

We adjust the step lengths $\gamma$ to give an acceptance rate during a pilot run of between $0.7$ and $0.8$ in the stationary phase. We obtain $\gamma=\{0.04,0.065,1.0\}$ for PMH$\{0,1,2\}$, respectively. Note that a single step length is used for each proposal to simplify the tuning. Of course, different step lengths can be used for each parameter, and we could also use different step lengths during the burn-in and the stationary phase of the algorithm using the approach discussed in Section~\ref{sec:constructPMH2}. As previously mentioned, the PMH2 algorithm avoids this (potentially difficult and time-consuming) procedure, by taking the local geometric information into account.

In the left column of Figure~\ref{fig:lgss-scaleinvariance}, we present the first $50$ iterations of the Markov chain from the three different algorithms. We note that the added information in the proposals of PMH1 and PMH2 aids the Markov chain in the burn-in phase. This results in that the Markov chains for the proposed algorithms reach the mode of the posterior quicker than the random walk used in PMH0.

To illustrate the scale invariance of the PMH2 algorithm, we reparametrise the LGSS model by $\{\theta_3,\theta_4\}=\{\phi,\sigma_v/10\}$. We keep the same settings as for the previous parametrisation and rerun the algorithms. From this run we obtain the middle column in Figure~\ref{fig:lgss-scaleinvariance}. We see clearly that the PHM1-algorithm does not perform well and gets stuck at the initial parameter value. The reason is that the second component of the gradient is increased by a factor 10 for the rescaled model. Since we still use the same step length, this will cause the PMH1 algorithm to overshoot the region of high posterior probability when proposing new values, and these will therefore never be accepted.

Finally, to improve the performance we recalibrate the three algorithms on the new parametrisation using the same procedure as before. We then obtain the new step lengths $\{0.005,0.0075,1.0\}$. The resulting Markov chains are presented in the right column of Figure~\ref{fig:lgss-scaleinvariance}. Despite the new step lengths, PMH0 and PMH1 continue to struggle. The reason is that the step lengths are limited by the small posterior variance in the $\theta_4$-parameter, resulting in a very slow progression in the $\theta_3$-direction. Again, for PMH2, the added Hessian information is used to rescale the proposal in each dimension resulting in a more efficient exploration of the posterior than for PMH0 and PMH1.

\subsection{The mixing of the Markov chains at stationarity}
We continue by investigating the mixing of the Markov chains at stationarity using an estimate of the integrated autocorrelation time (IACT) given by
\begin{align}
	\widehat{\textsf{IACT}}(\theta_{1:M}) = 1 + 2 \sum_{k=1}^{K} \widehat{\rho}_k (\theta_{1:M}),
\end{align}
where $\widehat{\rho}_k(\theta_{1:M})$ denotes the empirical autocorrelation at lag $k$ of $\theta_{1:M}$ (after the burn-in has been discarded). A low value of the IACT indicates that we obtain many uncorrelated samples from the target distribution, implying that the chain is mixing well. Here, $K$ is determined as the first index for which the empirical autocorrelation satisfies $|\widehat{\rho}_K(\theta_{1:M})| < 2/\sqrt{M}$, i.e.\ when the coefficient is statistically insignificant.

We return to the LGSS model in \eqref{eq:lgss} with the original parameterisation $\{\theta_1,\theta_2\}=\{\phi,\sigma_v\}$ using the same settings as before. A total of $25$ data sets are generated using the parameters $\theta^{(1)}$ and the algorithms are initialised at the true parameter values to avoid a long burn-in phase. The step sizes are determined using a series of pilot runs on the first generated dataset to minimise the total IACT for each algorithm. This is done to make a fair comparison between the different algorithms at their near \textit{optimal} performance. The resulting step sizes are obtained as $\{0.08, 0.075, 1.50\}$. 

\begin{table}[p]
	\centering
	\begin{tabular}{llccccc}
		\toprule
		&& \textbf{Acc.\ rate} & \multicolumn{2}{c}{\textbf{IACT($\phi$)}} & \multicolumn{2}{c}{\textbf{IACT($\sigma_v$)}} \\
		 \cmidrule(r){3-3} \cmidrule(r){4-5} \cmidrule(r){6-7}
		 && Median & Median & IQR & Median & IQR \\
\midrule
\parbox[t]{2mm}{\multirow{6}{*}{\rotatebox[origin=c]{90}{\textsf{PMH0}}}}
& bPF(500)  & 0.02 & 257 & 146 & 265 & 371 \\ 
&bPF(1000) & 0.06 & 83  & 129 & 79 & 118 \\ 
&bPF(2000) & 0.15 & 29  & 23 & 15 & 24 \\ 
&faPF(50)  & 0.37 & 9   & 8 & 8 & 5 \\ 
&faPF(100) & 0.38 & 9   & 6 & 7 & 4 \\ 
&faPF(200) & 0.38 & 7   & 6 & 7 & 4 \\ 
\midrule
\parbox[t]{2mm}{\multirow{6}{*}{\rotatebox[origin=c]{90}{\textcolor{darkred}{\textsf{PMH1}}}}}
&bPF(500) & 0.02 & 187 & 271 & 203 & 347 \\ 
&bPF(1000) & 0.10 & 64 & 85 & 49 & 72 \\ 
&bPF(2000) & 0.22 & 23 & 16 & 12 & 24 \\ 
&faPF(50) & 0.58 & \textbf{3} & 2 & \textbf{3} & 1 \\ 
&faPF(100) & 0.59 & 4 & 2 & \textbf{3} & 1 \\ 
&faPF(200) & 0.58 & \textbf{3} & 1 & \textbf{3} & 1 \\ 
\midrule
\parbox[t]{2mm}{\multirow{6}{*}{\rotatebox[origin=c]{90}{\textcolor{steelblue}{\textsf{PMH2}}}}}
&bPF(500) & 0.03 & 170 & 211 & 164 &190 \\ 
&bPF(1000) & 0.10 & 59 & 73 & 65 &80 \\ 
&bPF(2000) & 0.24 & 13 & 10 & 19 &17 \\ 
&faPF(50) & 0.66 & \textbf{3} & 1 & 4 & 2 \\ 
&faPF(100) & 0.66 & \textbf{3} & 1 & 5 & 2 \\ 
&faPF(200) & 0.66 & \textbf{3} & 1 & 4 & 2 \\ 
\bottomrule
\end{tabular}
\caption{Median and IQR for the acceptance rate and IACT using different SMC algorithms. The values are computed using $25$ different data sets from the LGSS model.}
\label{tbl:pmh-lgss}
\end{table}

Finally, we estimate the mixing in each of the $25$ simulated data sets during $M=30 \thinspace 000$ MCMC iterations (discarding the first $10 \thinspace 000$ iterations as burn-in). The results are presented in Table~\ref{tbl:pmh-lgss}, where the median and interquartile range (IQR; the distance between the $25\%$ and $75\%$ quartiles) are presented for each PMH algorithm. Here, we present the results the standard version of Algorithm~\ref{alg:SMCfull}.

We see that the added information decreases the IACT about $2$ times for PMH1 and PMH2 compared with PMH0. We conclude that the extra information brought by the gradient and the Hessian improves the mixing of the Markov chains in this model, which results in a more efficient exploration of the posterior. Note that, for this parametrisation of the LGSS model the posterior is quite isotropic (which can also be seen in the left column of Figure~\ref{fig:lgss-scaleinvariance}). Hence, the conditions are in fact rather favourable for PMH0 and PMH1.

\subsection{Parameter inference in a Poisson count model}
\label{sec:results:earth}
In this section, we analyse the annual number of major earthquakes\footnote{The data is obtained from the Earthquake Data Base System of the U.S. Geological Survey, which can be accessed at \url{http://earthquake.usgs.gov/earthquakes/eqarchives/}.} (over $7$ on the Richter scale) during the period from year $1900$ to $2014$. Following \citet{Langrock2011}, we model the data using
\begin{subequations}
\begin{align}
	x_{t+1}|x_{t} &\sim \mathsf{N} \Big( x_{t+1}; \phi x_t, \sigma^2 \Big), \\
	y_{t}|x_{t}   &\sim \mathsf{P} \Big( y_t; \beta \exp(x_t) \Big),
\end{align}
\label{eq:earthquakemodel}
\end{subequations}
with parameters $\theta=\{\phi,\sigma,\beta\}$ and uniform priors over $|\phi| < 1$, $\sigma > 0$ and $\beta >0$. Here, $\mathsf{P}(\lambda)$ denotes a Poisson distribution with parameter $\lambda$.

We repeat the procedure from the previous subsection and obtain the step lengths $\{0.06,0.006,0.85\}$. Here, we use $M = 30 \thinspace 000$ MCMC iterations (discarding the first $10 \thinspace 000$ iterations as burn-in), the bPF with systematic resampling, $\Delta=12$, $\theta_0=\{0.5, 0.5, 18\}$ and $L=2 \thinspace 500$. In this model, the estimate of the negative Hessian is often non--PD (during about half of the iterations) and the choice of regularisation is therefore important. To explore the properties of the regularisation, we apply both the standard and hybrid version of the PMH2 algorithm discussed in Section~\ref{sec:regularisation}. We compare these methods to standard and pre-conditioned versions of the the PMH0 and PMH1 algorithms, using the sample posterior covariance matrix calculated in the same manner as for the hybrid PMH2 algorithm.

\begin{sidewaystable}[p]
	\centering
	\begin{tabular}{lllccccccc}
		\toprule
		 & Version & SMC alg.\ & \textbf{Acc.\ rate} & \multicolumn{2}{c}{\textbf{IACT($\phi$)}} & \multicolumn{2}{c}{\textbf{IACT($\sigma$)}} & \multicolumn{2}{c}{\textbf{IACT($\beta$)}} \\
		 \cmidrule(r){4-4} \cmidrule(r){5-6} \cmidrule(r){7-8} \cmidrule(r){9-10}
		 &&& Median & Median & IQR & Median & IQR & Median & IQR \\
\midrule
\parbox[t]{2mm}{\multirow{4}{*}{\rotatebox[origin=c]{90}{\textsf{PMH0}}}}
&Standard&bPF(500) & 0.26 & 497 & 712 & 16 & 3 & 2639 & 1163 \\ 
&Standard&bPF(1000) & 0.30 & 89 & 150 & 15 & 3 & 2680 & 438 \\ 
&Pre-cond. & bPF(500) & 0.43 & 35 & 17 & 16 & 1 & 107 & 105 \\ 
&Pre-cond. & bPF(1000) & 0.45 & 38 & 28 & 16 & 2 & 129 & 131 \\ 
\midrule
\parbox[t]{2mm}{\multirow{4}{*}{\rotatebox[origin=c]{90}{\textcolor{darkred}{\textsf{PMH1}}}}}
 &Standard&bPF(500) & 0.76 & 665 & 442 & 277 & 162 & 2651 & 364 \\ 
 &Standard&bPF(1000) & 0.82 & 490 & 134 & 205 & 30 & 2875 & 1007 \\ 
 & Pre-cond. & bPF(500) & 0.62 & 266 & 187 & \textbf{9} & 3 & 1728 & 1638 \\
 & Pre-cond. & bPF(1000) & 0.70 & 98 & 209 & \textbf{9} & 3 & 1480 & 1732 \\
\midrule
\parbox[t]{2mm}{\multirow{4}{*}{\rotatebox[origin=c]{90}{\textcolor{steelblue}{\textsf{PMH2}}}}}
 &Standard & bPF(500) & 0.24 & 91 & 17 & 53 & 14 & 222 & 37 \\ 
 &Standard & bPF(1000) & 0.28 & 60 & 14 & 47 & 17 & 139 & 59 \\ 
 &Hybrid & bPF(500) & 0.45 & 20 & 3 & 17 & 4 & 30 & 15 \\ 
 &Hybrid & bPF(1000) & 0.49 & \textbf{17} & 4 & 18 & 3 & \textbf{23} & 5 \\ 
\bottomrule
\end{tabular}
\caption{Median and IQR for the acceptance rate and IACT using different number of particles. The values are computed using $10$ runs on the Earthquake count data model.}
\label{tbl:pmh-earth}
\end{sidewaystable}

In Table~\ref{tbl:pmh-earth}, we present the resulting acceptance rates and IACT values for each parameter and algorithm. We note the large decrease in IACT for $\beta$ when using the Hessian information, where the hybrid PMH2 seems to perform better than standard version for this model. The improved mixing by using PMH2 is due to the scale invariance property, as the parameter $\beta$ is at least an order of magnitude larger than $\phi$ and $\sigma$ (c.f.\ Figure~\ref{fig:lgss-scaleinvariance}). Note that a reparameterisation or using separate step lengths for the parameters could possibly have helped in improving the mixing in $\beta$ for the standard versions of PMH0 and PMH1.

Using the standard and hybrid version of PMH2, decreases the overall computational cost by a factor of about $100$ for a specific number of effective samples. The poor performance of the pre-conditioned algorithms is probably due to that the sample posterior covariance matrix does not fully capture the geometry of the posterior distribution.

\begin{figure}[p]
	\centering
	\includegraphics[width=\columnwidth]{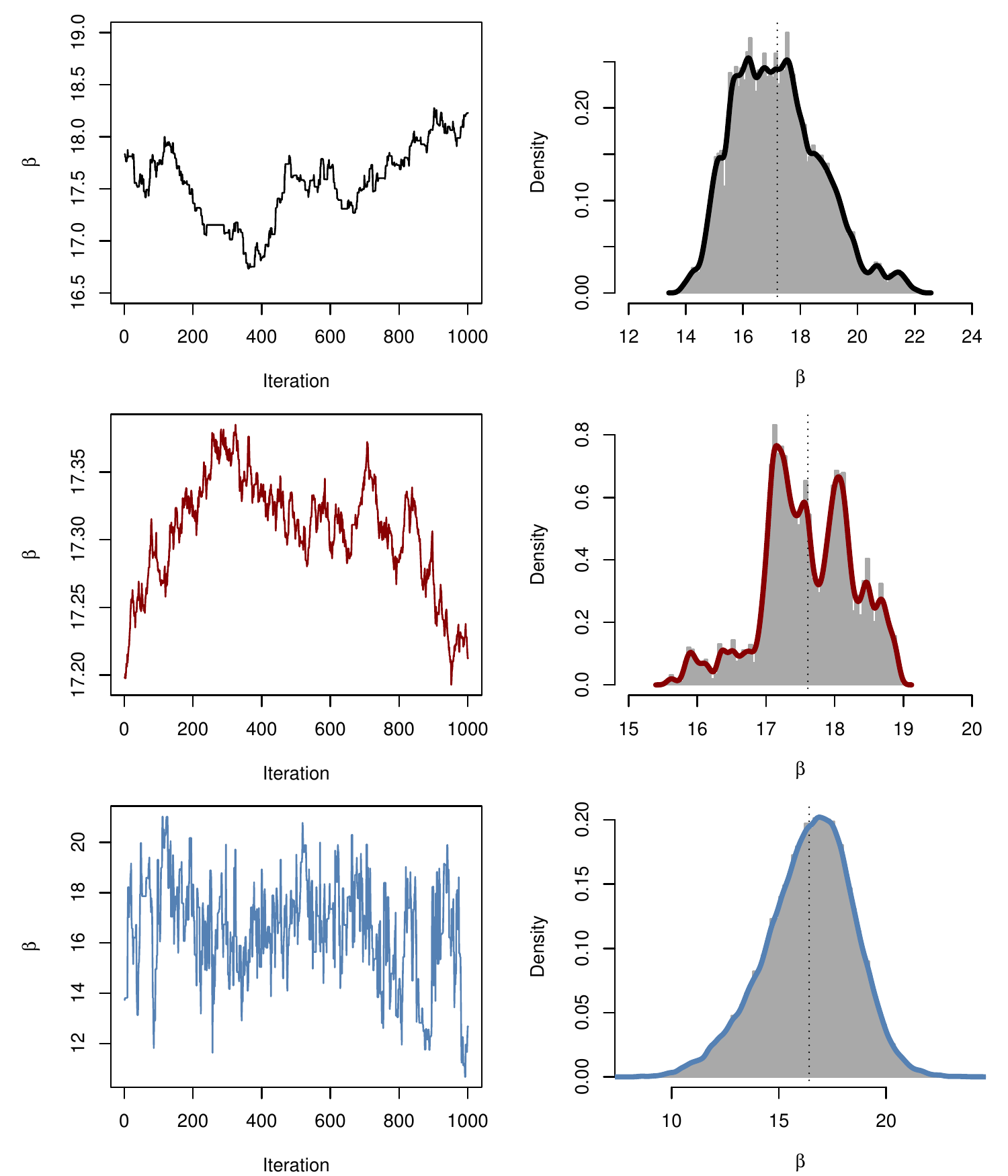}
	\caption{Part of the trace (left) and posterior estimates (right) for the $\beta$ parameter in the earthquake count model using standard versions of PMH0 (black), PMH1 (red) and hybrid version of PMH2 (blue). Dotted lines indicate the posterior means.}
	\label{fig:earthquake-beta-post}
\end{figure}

In Figure~\ref{fig:earthquake-beta-post}, we present the trace and posterior estimates for $\beta$ using the standard versions of PMH0 and PMH1 as well as hybrid PMH2. The posterior estimates are obtained by pooling the $10$ parallel Markov chains after the burn-ins have been discarded. We see that the traces behave rather differently with hybrid PMH2 exploring the space well compared with the other methods. 

Using the parameter posterior estimate, we can compute point estimates for the parameters of the model. The posterior mean for hybrid PMH2 is obtained as $\{0.88, 0.15 , 16.58\}$ with standard deviations $\{0.07,0.03,2\}$. The parameter estimate is comparable to the estimate $\{0.88,0.15,17.65\}$ obtained by a maximum likelihood-based method using the same data and model in \citet[Example 4.9]{Dahlin2014}. 

\subsection{Robustness in the lag and step size}
The PMH2 algorithm requires a number of parameters to be select by the user for each parameter inference problem. It is therefore interesting to discuss the robustness of the method with respect to these parameters. In the previous illustrations, we have seen that the number of particles $N$ is an important factor in determining the mixing. 

\begin{figure}[p]
	\centering
	\includegraphics[width=\columnwidth]{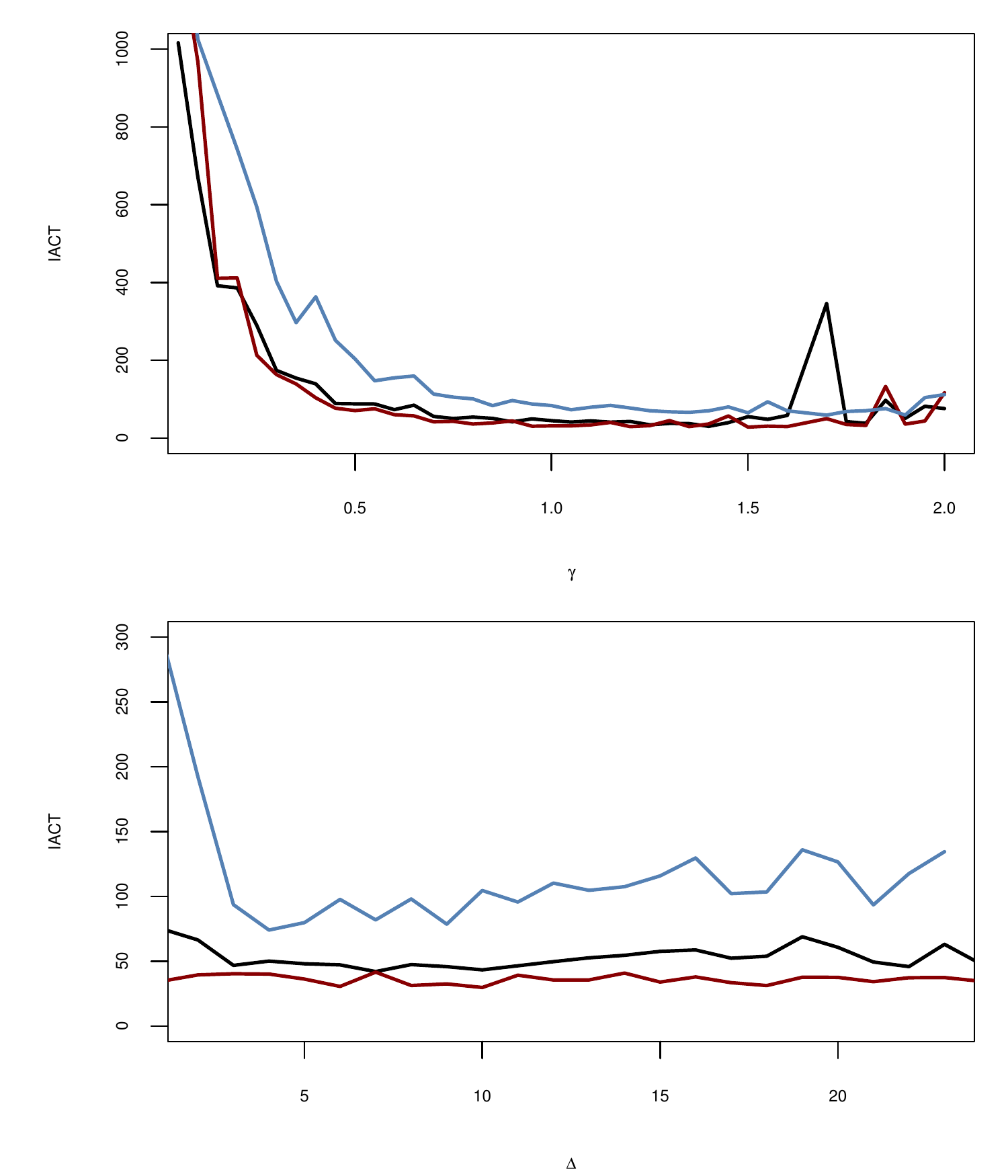}
	\caption{The IACT for $\phi$ (black), $\sigma$ (red) and $\beta$ (blue) for varying step sizes $\gamma$ (upper) and lag $\Delta$ (lower). The values are computed as the median of $10$ runs using standard PMH2 with the same data.}
	\label{fig:earthquake-sensitivity}
\end{figure}

Two other important parameters are the step length $\gamma$ and the lag in the FL-smoother $\Delta$. To illustrate the impact of these quantities on the IACT, we return to the Earthquake model in \eqref{eq:earthquakemodel} using the standard PMH2 algorithm with the same settings but with $M= 15 \thinspace 000$ (discarding the first $5 \thinspace 000$ iterations as burn-in) and $N=1 \thinspace 500$. In Figure~\ref{fig:earthquake-sensitivity}, we present the IACT for the three parameters in the model when varying $\gamma$ and $\Delta$, keeping everything else fixed. The standard PMH2 algorithm seems to be rather robust to both the choice of $\Delta$ and $\gamma$ after a certain threshold. Recall the discussion in Section~\ref{sec:results:flsmoother} for the FL smoother. We conclude that a suitable standard choice for the step length could be $\gamma=1$, which can be fine tuned if the performance is not good enough. This recommendation is also common in the literature concerning Newton-type algorithms.


\section{Discussion and future work}
Adding the gradient and Hessian information to the PMH proposal can have beneficial results including: (i) a shorter burn-in phase, (ii) a better mixing of the Markov chain, and (iii) scale-invariance of the proposal which simplifies tuning. The latter point is true in particular for PMH2, since this method takes the local curvature of the posterior into account, effectively making the method invariant to affine transformations.

It is common to distinguish between two phases of MCMC algorithms: the burn-in and stationary phases. We have seen empirically that the proposed methods can improve upon the original PMH0 during both of these phases but the \textit{best} choices for the step lengths can differ between these two phases. Typically, a smaller step length is preferred during burn-in and a larger during stationarity (the opposite holds for PMH0). The reason for this is that during burn-in, the (natural) gradient information will heavily skew the proposal in a direction of increasing posterior probability. That is, the methods tend to be \textit{aggressive} and propose large steps to make rapid progression toward regions of high posterior probability. While this is intuitively appealing, the problem is that we require the Markov chains to be reversible at all times. The reverse of these large steps can have very low probability which prevents them from being accepted.

One interesting direction for future work is therefore to pursue adaptive algorithms (see e.g.\ \citet{AndrieuThoms2008,PetersHosackHayes2010,PittSilvaGiordaniKohn2012}), to automatically tune the step lengths during the different phases of the algorithms. 
Another interesting possibility is to relax the reversibility requirement during burn-in; see \citep{DiaconisHolmesNeal2000} for a related reference. This would cause the methods to behave like optimisation procedures during the initial phase, but transition into samplers during the second phase.

Finally, another very interesting direction for future work is to extend the proposed methods to develop a particle-version of the manifold Hamiltonian Monte Carlo (mHMC) algorithm \citep{DuaneKennedyPendletonRoweth1987,Neal2010,GirolamiCalderhead2011}. The reason for this is motivated by the large improvement in mixing seen by e.g.\ \citet{Neal2010,GirolamiCalderhead2011} for high dimensional problems in \textit{vanilla} MH sampling.

\bibliographystyle{plainnat}
\bibliography{dahlin}
\end{document}